\documentclass[12pt]{article}
\usepackage{amsmath, amsfonts, amssymb, amsthm, txfonts, enumerate, graphicx}
\usepackage{epstopdf}

\def\SSI{{\rm \bf SSI}}
\def\BI{{\rm \bf BI}}

\renewcommand{\baselinestretch}{1.5}

\begin{document}
\renewcommand{\baselinestretch}{1.5}
\begin{center}
{\large\bf  A Three-Dimensional Voting System in Hong Kong}\\
{\sc Wai-Shun Cheung${}^{a}$ and Tuen-Wai Ng${}^{b,*}$}\\
{\it ${}^{a,b}$Department of Mathematics,\\
The University of Hong Kong,\\
Pokfulam, Hong Kong.\\}
\quad E-mail address:  ${}^{a}$cheungwaishun@gmail.com,\, ${}^{b}$ntw@maths.hku.hk
\end{center}
\begin{abstract}
The voting system in the Legislative Council of Hong Kong (Legco) is sometimes unicameral and sometimes bicameral, depending on whether the bill is proposed by the Hong Kong government.  Therefore, although without any representative within Legco,  the Hong Kong government has certain degree of legislative power --- as if there is a virtual representative of the Hong Kong government within the Legco. By introducing such a virtual representative of the Hong Kong government, we show that Legco is a three-dimensional voting system. We also calculate two power indices of the Hong Kong government through this virtual representative and consider the $C$-dimension and the $W$-dimension of Legco. Finally, some implications of this Legco model to the current constitutional reform in Hong Kong will be given. 
\end{abstract}

\begin{figure}[b]  
\rule[-2.5truemm]{5cm}{0.1truemm}\\[2mm] 
{ \footnotesize 
{\rm *Corresponding author}.\, Department of Mathematics, The University of Hong Kong, Pokfulam, Hong Kong. Tel:(852)22415631; fax (852) 25592225. E-mail address: ntw@maths.hku.hk (T.W. Ng).
}
 
 \end{figure}

\newpage
\bigskip
\noindent
\section{Introduction}
\medskip
The voting system used in the Hong Kong Legislative Council (Legco) is unique within the
current, global range of electoral systems because it is sometimes unicameral and sometimes bicameral, depending on whether the bill is proposed by the government of the Hong Kong Special Administrative Region (HKSAR). Because of this special feature, it would be interesting to study this voting system from the mathematical point of view. In this article, we shall construct a mathematical model of this voting system and conduct a detailed mathematical analysis of it. For example, we shall measure the complexity of the voting mechanism of Legco by computing its dimension.
 
It was proposed in Taylor (1995) that different voting systems can be classified by their dimensions. We shall state the precise definition of the dimension of a voting system in the next section but it is easier to grasp the idea by considering some examples. The usual unicameral system in which the bills will be passed by a simple majority vote of the members is a one dimensional voting system. Bicameral legislatures require a concurrent majority to pass legislation are two dimensional voting systems. In 1995, Alan D. Taylor mentioned in Taylor (1995) that he did not know any real-world voting system of dimension $3$ or higher and this point was reiterated in page 255 of the second edition of Taylor (1995). Such a real-world  voting system of dimension $3$ was first provided by Josep Freixas in 2004. He showed in Freixas (2004) that the dimension of the European Union Council under the Nice rules (since 2000) is $3$. In this article, we shall show that the voting system of Legco (since 1998) is also of dimension $3$. So we have two very different real voting systems of dimension $3$.

By 2014, Legco consists of two groups of legislators: one group comprising $35$ members selected in the functional constituencies and the other group comprising the remaining $35$ members elected by universal suffrage in the geographical constituencies. About half of the $35$ functional constituencies seats go to business sectors, about one-third belongs to sectors for professionals and the rest are for representatives of social organizations or district councils. For a detailed analysis of the functional constituencies, see Loh et al. (2006) and Ma (2009).

Since the handover of Hong Kong in 1997, in order to strengthen the executive dominance over the legislature, the Basic Law (the constitutional document for the HKSAR) requires the passage of motions, bill or amendments to government bills introduced by legislators to pass by the concurrent majorities of two groups. On the other hand, motions, amendments to motions, bills and amendments to bill raised by the Hong Kong government only need a simple majority vote of the members present to pass.

Therefore, Legco is sometimes unicameral and sometimes bicameral, depending on whether the bill is proposed by the Hong Kong government. This unique feature of Legco makes the computation of the dimension of it a non-trivial task. For example, given a coalition of $35$ members from the geographical constituencies and $2$ members from the functional constituencies, there is no way to tell if it is a winning coalition or a losing coalition (unless you know if the bill is proposed by the Hong Kong government). To overcome this difficulty, we introduce a virtual representative of the Hong Kong government, which will vote for a bill if it is proposed by the government but against a bill otherwise.
 To be more precise, the $35$ members from geographical constituencies are numbered $1,...,35$, the $35$ members from functional constituencies are numbered $36,...,70$,  and the virtual member (the government) is numbered $71$.  Then, a coalition or subset $S$ of the set $\{1,2,...,70,71\}$ is a winning coalition if and only if 

(a) $71 \in S$ and $|S\cap\{1,...,70\}| \ge 36$, or

(b) $71 \notin S$, $|S \cap \{1,...,35\}| > 18$ and $|S \cap \{36,...,70\}| > 18$.\\
\noindent
where $V \cap U$ is the intersection of the two coalitions $V$ and $U$ and $|U|$ is the number of members in the coalition $U$.

 With this mathematical model of Legco, the voting system of Legco becomes a simple game (see the definition in the next section) and it is now possible to compute the dimension of Legco which will turn out to be $3$ (see section 3).

Although there is no representative of the government of HKSAR in Legco, this voting system has given the Hong Kong government a certain degree of legislative power. With the introduction of member $71$, the virtual representative of the government, we can then calculate various power indices of the Hong Kong government within Legco and compare its power with that of any individual Legco member. This will be done in section 4. Finally, in section 5, we shall discuss some implications of our mathematical analysis of Legco to the current constitutional reform in Hong Kong.

\section{Dimension of a simple game}
\medskip

\noindent
{\bf Definition 1.} A (monotonic) {\it simple} game or a voting system is a pair $(N,v)$ where $N=\{1,\ldots,n\}$ is the set of players and $v:2^N\rightarrow\{0,1\}$ is the characteristic function defined on the power set $2^N$ of $N$, which satisfies $v(\phi)=0$, $v(N)=1$ and $v(S)\le v(T)$ whenever $S\subseteq T$. A coalition of players $S\subseteq N$ is {\it winning} if $v(S)=1$ and {\it losing} if $v(S)=0$.

A {\it weighted majority game} is a simple game which can be realized by a vector $(w_1 , . . . , w_n )$ together with a threshold $q$ which makes
the representation $[q; w_1,\ldots,w_n]$ in such a way that $S$ is a winning coalition if and only if $\sum_{j\in S}w_j\ge q$.

A {\it weighted $m$-majority game} is a simple game which can be expressed by $m$ realizations $[q^i; w_1^i,\ldots,w_n^i]$, $1\le i\le m$, in such a way that $S$ is a winning coalition if and only if $\sum_{j\in S}w_j^i\ge q^i$ for all $1\le i\le m$.   So a weighted $m$-majority game can be considered as an intersection of $m$ weighted majority games and we can also represent it by an amalgamated matrix 

$$\begin{pmatrix}q^1;& w_1^1& \cdots & w_n^1 \\ \vdots &&& \vdots \\ q^m;& w_1^m &\cdots& w_n^m\end{pmatrix}.$$ \\

Every weighted majority game is obviously a (monotonic) simple game. However, the converse is in general not true and we shall see in the next section that the voting system of Legco is not a weighted majority game. To show that certain voting system cannot be realized as a weighted majority game, it would be useful to know that any weighted majority game must be {\it swap robust}, namely, for any two winning coalitions $S$ and $S'$  in a weighted voting system, if we make a one-for-one exchange of players, then at least one of the two resulting coalitions must still be a winning coalition (see Taylor (1995)). Here one of the players in the swap must belong to S but not $S'$, and the other must belong to $S'$ but not $S$.

Even though not any simple game is a weighted majority game, it was proved in Taylor (1995) that every simple game can be realized as 
a weighted $m$-majority game.  The smallest such possible $m$ is called the {\it dimension} of the game.

The computation of the dimension of
a simple game has been proved to be a hard computational problem (see
Deineko and Woeginger (2006)), thus our calculation is a complex task which
will be done by combinatorial arguments for the particular voting system at
hand. The dimension of other complex voting systems using combinatorial
arguments were studied quite recently in Freixas and Puente (2008).

However, even if two games are of the same dimension, they may not be equivalent. In the literature, different types of dimensions have been introduced (see Freixas and Marciniak (2009)).  Before giving their definitions, we need to introduce two orderings related to the power of individual members in a game (see Carreras and Freixas (2008)).

For any $i,j\in N$, we say that $i\le_D j$ if for any $U\subset N$ such that $i,j\notin U$, we have $U\cup\{j\}$ is a winning coalition whenever $U\cup\{i\}$ is.  A game is {\it complete} if $\le_D$ is total.  It is known that a game is complete if and only if it is swap robust.

An $i\in N$ is {\it crucial} in a winning coalition $U$ if $U\setminus \{i\}$ is no longer winning.  We say that $i\le_d j$ if the number of winning coalitions of size $k$ containing $i$ with $i$ crucial is smaller than or equal to the number of those containing $j$ with $j$ crucial for all $1\le k\le n$. A game is {\it weakly complete} if $\le_d$ is total.

Every simple game can be realized as an interaction of a finite number of  complete games, the smallest  possible number is called the {\it C-dimension} of the game.  Likewise, every simple game can be realized as an interaction of a finite number of weakly complete games, the smallest possible number is called the {\it W-dimension} of the game.  For a detailed analysis of various types of dimensions, we refer the readers to  Freixas and Marciniak (2009).

Every weighted game is complete and every complete game is weakly complete (see Carreras and Freixas (2008)). Hence,
\begin{enumerate}
\item $W$-dimension($v$) $\le$ $C$-dimension($v$) $\le$ dimension($v$) for all simple game $v$;

\item  if the simple game $v$ is weighted, then dimension($v$) is $1$ and hence both $W$-dimension($v$) and $C$-dimension($v$) are equal to $1$;

\item if the simple game $v$ is complete, then $C$-dimension($v$) is equal to $1$ and therefore $W$-dimension($v$) is also equal to $1$.
\end{enumerate}

\section{Hong Kong Legco System}
\medskip
Recall that the Legco members are divided into two groups: half of the members are returned by geographical constituencies through direct elections, and the other half by functional constituencies. On the writing of this article, there are $70$ members in the current fifth term Legco (2012-2016). The composition of Legco beyond the third term is not specified in the Basic Law (the constitutional document for the Hong Kong Special Administrative Region). However, article 68 of the Basic Law requires that 

{\it ``The method for forming the Legislative Council shall be specified in the light of the actual situation in the Hong Kong Special Administrative Region and in accordance with the principle of gradual and orderly progress. The ultimate aim is the election of all the members of the Legislative Council by universal suffrage.''}

While the Basic Law now no longer expressly dictates the formation of the fifth term Legco of HKSAR in the year 2012, in December 2007, the Standing Committee of National People's Congress (SCNPC) decided  that

{\it ``The ratio of functional constituency members to geographical constituency members shall not be changed and the procedures for voting on bills and motions in the Legislative Council shall remain unchanged.''}
 
The details of the decisions of SCNPC in December 2007 can be found in http://www.legco.hk (the official website of Hong Kong Legislative Council).  

In view of the above constraints on the pace of constitutional reform in Hong Kong, it makes sense to assume that Legco has $2n$ legislative members, say represented by $1,...,2n$, with $n$ an arbitrary positive integer. We further assume that among the $2n$ Legco members, the first $n$ members are returned by geographical constituencies through direct elections, while the remaining $n$ members are returned by functional constituencies. 

We also assume that a bill will be passed if 
\begin{enumerate}
\item[(a)] the bill is proposed by the Hong Kong government and it is supported by a simple majority in Legco; or
\item[(b)] the bill is proposed by a Legco member and it is supported by a simple majority in each of the two groups.
\end{enumerate}

Finally, we again introduce a virtual representative of the Hong Kong government (denoted by $2n+1$), which will vote for a bill if it is proposed by the government but against a bill otherwise.  Now we can reformulate Legco as a simple game of size $2n+1$.\\

\noindent
{\bf Definition 2}. Legco of size $2n$ is a (monotonic) simple game of $2n+1$ players $1,\ldots,2n+1$,  such that $S$ is a winning coalition if one of the following holds:
\begin{enumerate}
\item[(a)] $2n+1\in S$ and $\left|S\cap \{1,\ldots, 2n\}\right| \ge n+1$
\item[(b)] $2n+1\notin S$ and $|S\cap \{1,\ldots, n\}| > \frac{n}{2}$ and $|S\cap \{n+1,\ldots, 2n\}| > \frac{n}{2}$.
\end{enumerate}

Our first result is a comparison of powers among individuals.

\noindent
{\bf Proposition 1.}
\begin{enumerate}[(a)]
\item $j=_D k$ for any $1\le j,k\le n$ or $n+1\le j,k\le 2n$.
\item $1=_d n+1$ but $1$ and $n+1$ are not $D$-comparable.
\item When $n$ is even, $j<_D 2n+1$ for any $1\le j\le 2n$.
\item If $n$ is odd and $n>4$, then $j$ and $2n+1$ are not $d$-comparable for any $1\le j\le 2n$.
\end{enumerate}

\noindent
{\bf Proof.}\, 
(a) and (b) are trivial.

For (c) and (d), we assume $j=1$. Consider $U=\{n+1,\ldots,2n\}$. Obviously (P): $U\cup\{2n+1\}$ is winning but $U\cup\{1\}$ is not.

(c) Suppose $n$ is even.  If $S\cup\{1\}$ is winning for some coalition $S\notni 1$, then

$|S\cap \{1,\ldots, n\}| \ge \frac{n}{2}$ and $|S\cap \{n+1,\ldots, 2n-1\}| \ge \frac{n}{2}+1.$

Thus $|S|\ge n+1$ and hence $S\cup\{2n+1\}$ is a winning coalition. Therefore $1\le_D 2n+1$.

Together with (P), we have $1<_D 2n+1$.

(d) Suppose $n$ is odd.  We have $S=\{1,\ldots, (n+1)/2\}\cup\{n+1, \ldots, (3n+1)/2\}$ which is a winning coalition of size $n+1$ with $1$ crucial, but there is no winning coalition of size $n+1$ with $2n+1$ crucial.  On the other hand, from (P), clearly there are less winning coalitions of size $2n$ with $1$ crucial then those with $2n+1$ crucial. Therefore, we have $1$ and $2n+1$ are not $d$-comparable.  $\hfill\blacksquare$

Our first main result is the following\\

\noindent
{\bf Theorem 1.} {\it Legco of size $2n$ is of dimension $1$ if $n=1$ or $2$, of dimension $2$ if $n=3$ or $4$, and of dimension $3$  if $n \ge 5$.}\\

\noindent
{\bf Proof.}\, For $n=1$ or $2$, Legco is of dimension $1$ as one can check easily that Legco of size $2$ is realized by $[2;1,1,0]$ and Legco of size $4$ is realized by $[4;1,1,1,1,1]$.

So we shall let $n \ge 3$ and we shall use $\lfloor x \rfloor$ and $\lceil x \rceil$ to denote the greatest integer less than or equal to $x$ and the least integer greater than or equal to $x$ respectively.

Recall that Legco is of dimension $1$ means that it is a weighted majority game (which must be swap robust). Let $S=\{1,...,\lfloor n/2 \rfloor,\lfloor n/2 \rfloor +1,n+1,...,n+\lfloor n/2 \rfloor,n+\lfloor n/2 \rfloor +1\}$ and $S'=\{1,...,\lfloor n/2 \rfloor,\lfloor n/2 \rfloor +2,n+1,...,n+\lfloor n/2 \rfloor,n+\lfloor n/2 \rfloor +2\}$. Then both of them are winning coalitions of Legco. Now if we swap $\lfloor n/2 \rfloor +1$ in $S$ with $n+\lfloor n/2 \rfloor +2$ in $S'$, then we get the following two coalitions

$$\{1,...,\lfloor n/2 \rfloor,n+1,...,n+\lfloor n/2 \rfloor,n+\lfloor n/2 \rfloor +1,\lfloor n/2 \rfloor +2\}$$ and $$\{1,...,\lfloor n/2 \rfloor,\lfloor n/2 \rfloor +1,\lfloor n/2 \rfloor +2,n+1,...,n+\lfloor n/2 \rfloor\}.$$

Since both of them are losing coalitions, Legco is not swap robust. Hence, for $n \ge 3$, Legco of size $2n$ cannot be of $C$-dimension $1$ and of dimension $1$ as it is not swap robust.

For $n=3$ or $4$, Legco is of dimension $2$ as one can check easily that Legco of size $6$ can be realized by $\begin{pmatrix}10;2,2,2,3,3,3,1\\10;3,3,3,2,2,2,1\end{pmatrix}$ and Legco of size $8$ can be realized by $\begin{pmatrix}15;2,2,2,2,3,3,3,3,4\\15;3,3,3,3,2,2,2,2,4\end{pmatrix}$.

From now on, we will assume that $n \ge 5$. We first show that Legco of size $2n$ can be realized as a $3$-weighted majority game.

Let 
$$A=\begin{pmatrix} n+1; & \overbrace{1 \cdots 1}^{n} & \overbrace{1 \cdots 1}^{n} & 0 \\ \displaystyle\frac{n+1}{2}; & 1 \cdots 1 & 0 \cdots 0 & \displaystyle\frac{n}{2} \\   \displaystyle\frac{n+1}{2}; & 0 \cdots 0 & 1 \cdots 1 & \displaystyle\frac{n}{2} \end{pmatrix}.$$
We will show that the winning coalitions of the game realized by $A$ are exactly the winning coalitions of Legco. Hence
$A$ realizes Legco and the dimension of Legco is less than or equal to $3$.

Note that $S \subset \{1,...,2n+1\}$ is a winning coalition of the $3$-weighted majority game defined by the amalgamated matrix $A$ if and only if 

\begin{enumerate}
\item[(i)] $\left|S\cap \{1,\ldots, 2n\}\right| \ge n+1$; and
\item[(ii)] $|S\cap \{1,\ldots, n\}| + \frac{n}{2}|S\cap \{2n+1\}| \ge \frac{n+1}{2}$; and
\item[(iii)]
$|S\cap \{n+1,\ldots, 2n\}| + \frac{n}{2}|S\cap \{2n+1\}| \ge \frac{n+1}{2}$.
\end{enumerate}

It is clear that $2n+1\in S$ and $i,ii,iii$ hold $\Leftrightarrow 2n+1\in S$ and $i$ holds.
Also, $2n+1\notin S$ and $i,ii,iii$ hold $\Leftrightarrow 2n+1\notin S$ and $i,ii$ hold. Therefore $S$ is a winning coalition of the game induced by the amalgamated matrix $A$ if and only if $S$ is a winning coalition of Legco and we are done.

Finally, it remains to show that the dimension of Legco cannot be $2$.

Suppose $B=\begin{pmatrix} q^1; & e_1^1 &\cdots &e_n^1 & f_1^1 &\cdots &f_n^1 & g^1 \\q^2; & e_1^2 &\cdots &e_n^2 & f_1^2& \cdots &f_n^2 & g^2 \end{pmatrix}$ realizes Legco. Let $W$ be a winning coalition of the game induced by $B$. If $W=E\cup F$, where $E\subseteq\{1,...,n\}$ and $F\subseteq\{n+1,...,2n\}$, then for any permutation $\sigma$ on $\{1,...,n\}$ and permutation $\tau$ on $\{n+1,..., 2n\}$, $\sigma(E)\cup \tau(F)$ is also a  winning coalition as any ordinary member of Legco has the same voting power.  Therefore, $\sum_{k\in E}e_{\sigma(k)}^i+\sum_{j\in F}f_{\tau(j)-n}^i\ge q^i$, for $i=1,2$.  

Start with an equality $\sum_{k\in E}e_{k}^i+\sum_{j\in F}f_{j}^i\ge q^i$ for a fixed $i$, by considering suitable permutation pairs $\sigma$ and $\tau$, we can obtain ${}_{n}C_{|E|} \times {}_{n}C_{|F|}$ similar inequalities. Summing all these inequalities and we notice that for each $k$ and $l$, there are ${}_{n-1}C_{|E|-1} \times {}_{n}C_{|F|}$\, $e_k^i$ and ${}_{n}C_{|E|} \times {}_{n-1}C_{|F|-1}$\, $f_l^i$ in the resulting inequality. Divide the whole inequality by ${}_nC_{|E|} \times {}_{n}C_{|F|}$, we have $|E|e^i+|F|f^i\ge q^i$, where $e^i=\frac1n(e_1^i+\cdots+ e_n^i)$ and $f^i=\frac1n(f_1^i+\cdot+f_n^i)$, $i=1,2$.  Similarly, if the winning coalition $W$ is of the form $W=E\cup F\cup\{2n+1\}$, where $E\subseteq\{1,...,n\}$ and $F\subseteq\{n+1,...,2n\}$, then we have $|E|e^i+|F|f^i + g^i\ge q^i$,$i=1,2$. 
Therefore, if we consider another amalgamated matrix $B'=\begin{pmatrix} q^1; & e^1 &\cdots &e^1 & f^1& \cdots &f^1 & g^1 \\q^2; & e^2 &\cdots &e^2 & f^2& \cdots &f^2 & g^2 \end{pmatrix}$, then any winning coalition of $B$ must be a winning coalition of $B'$. By using a similar argument, we can also show that any losing coalition of $B$ must be a losing coalition of $B'$. Hence, the games induced by $B$ and $B'$ will have exactly the same set of the winning coalitions.
Therefore without loss of generality, for $B$, we can assume that $e_1^i=\cdots=e_n^i=e^i$ and $f_1^i=\cdots=f_n^i=f^i$, $i=1,2$.
Since $e_k^i$ and $f_k^i$ are non-negative and $B$ realizes Legco, we must have $e^i>0$ and $f^i>0$.

Note that $\{1,\ldots, \lfloor n/2 \rfloor+1, n+1,\ldots, n+\lfloor n/2 \rfloor+1\}$ is a winning coalition of Legco and hence $(\lfloor n/2 \rfloor+1)e^i+(\lfloor n/2 \rfloor+1)f^i\ge q^i$, $i=1,2$.

On the other hand, $\{1,\ldots,n, n+1, \ldots, n+\lfloor n/2 \rfloor\}$ is not a winning coalition of Legco and hence either $ne^1+\lfloor n/2 \rfloor f^1< q^1$ or $ne^2+\lfloor n/2 \rfloor f^2< q^2$ .  If it is the former case, then $ne^1+\lfloor n/2 \rfloor f^1<  q^1\le (\lfloor n/2 \rfloor+1)e^1+(\lfloor n/2 \rfloor+1)f^1$ and thus $(\lceil n/2 \rceil-1)e^1<f^1$.  Similarly if it is the latter case, then  
$(\lceil n/2 \rceil-1)e^2<f^2$.

By symmetry, we also have $(\lceil n/2 \rceil-1)f^1<e^1$ or $(\lceil n/2 \rceil-1)f^2<e^2$.  Note that if $(\lceil n/2 \rceil-1)e^1<f^1$ and $(\lceil n/2 \rceil-1)f^1<e^1$, then we have $(\lceil n/2 \rceil-1)^2 < 1$ which is impossible for $n \ge 3$. Hence, without loss of generality, we may assume that $(\lceil n/2 \rceil-1)e^1<f^1$ and $(\lceil n/2 \rceil-1)f^2<e^2$.

Now $\{1, \ldots, \lceil n/2 \rceil+1, n+1, \ldots, n+\lfloor n/2 \rfloor, 2n+1 \}$(which has $n+1$ ordinary Legco members) is a winning coalition of Legco, thus $(\lceil n/2 \rceil+1)e^1+(\lfloor n/2 \rfloor)f^1+g^1\ge q^1$. It follows from $(\lceil n/2 \rceil-1)e^1<f^1$ that $2e^1+(\lfloor n/2 \rfloor+1)f^1+g^1\ge q^1$.

Clearly, $\{1, n+1, \ldots, 2n, 2n+1\}$ is also a winning coalition of Legco, thus $e^2+nf^2+g^2\ge q^2$. It follows from $(\lceil n/2 \rceil-1)f^2<e^2$ that $2e^2+(\lfloor n/2 \rfloor+1)f^2+g^2\ge q^2$.

Since the set $S=\{1, 2, n+1, \ldots, n+\lfloor n/2 \rfloor+1, 2n+1\}$ has $\lfloor n/2 \rfloor+3$ ordinary Legco members and $\lfloor n/2 \rfloor+3 < n+1$ for $n \ge 5$, $S$ is a losing coalition of Legco. On the other hand, we have  $2e^1+(\lfloor n/2 \rfloor+1)f^1+g^1\ge q^1$  and $2e^2+(\lfloor n/2 \rfloor+1)f^2+g^2\ge q^2$ and hence $S$ is a winning coalition of the 2-majority game induced by $B$. This is a contradiction and therefore $B$ cannot realize Legco and we are done.
$\hfill\blacksquare$

\medskip
By Theorem 1 and Theorem 2.1 of Freixas (2004), both Lego and European Union Council under the Nice rules (since 2000) are of dimension $3$.
However, they are very different voting systems. For example, the voting system of the European Union Council has a huge number of voters and the system itself is sophistically described with a great number of different equi-desirability classes, while the
Legco voting system has just three types of voters. On the other hand, the European voting system is a complete (or swap robust) simple game
while the Legco voting system is not by Proposition 1. Moreover, it was pointed out by the referee of this paper that the voting system of the European Union Council
has dimension $3$ but $C$-dimension $1$ (which implies that it also has $W$-dimension $1$) and suggested to find out the $C$-dimension and the $W$-dimension of Legco. \\   

\noindent
{\bf Theorem 2.}\begin{enumerate}
\item When $n$ is even and $n>4$,  Legco of size $2n$ is of $W$-dimension $1$ and $C$-dimension $2$.
\item When $n$ is odd and $n>4$, Legco of size $2n$ is of $W$-dimension $2$ and $C$-dimension $3$.
\end{enumerate}

\noindent
{\bf Proof.}\, 
\begin{enumerate}
\item When $n=2k$ is even, the voting system is weakly complete by Proposition $1$.  It is not swap robust and therefore must be of $C$-dimension greater than $1$.  Note that it is the intersection of the two compete systems  below:

System 1:
\begin{enumerate}
\item[(a)] $2n+1\in S$ and $\left|S\cap \{1,\ldots, 2n\}\right| \ge 2k+1$
\item[(b)] $2n+1\notin S$ and $\left|S\cap \{1,\ldots, 2n\}\right| \ge 2k+2$ and $|S\cap \{1,\ldots, n\}| \ge k+1$.
\end{enumerate}

System 2:
\begin{enumerate}
\item[(a)] $2n+1\in S$ and $\left|S\cap \{1,\ldots, 2n\}\right| \ge 2k+1$
\item[(b)] $2n+1\notin S$ and $\left|S\cap \{1,\ldots, 2n\}\right| \ge 2k+2$ and $|S\cap \{n+1,\ldots, 2n\}| \ge k+1$.
\end{enumerate}

Therefore it is of $C$-dimension $2$.

\item
When $n=2k+1$ is odd, the voting system is not weakly complete as $1$ and $2n+1$ are not $d$-comparable. However, it is the intersection of two weakly completed systems realized by the matrices below:

$$\begin{pmatrix} n+1; & \overbrace{1 \cdots 1}^{n} & \overbrace{1 \cdots 1}^{n} & 0 \end{pmatrix}.$$
$$\begin{pmatrix} \displaystyle\frac{n+1}{2}; & 1 \cdots 1 & 0 \cdots 0 & \displaystyle\frac{n}{2} \\   \displaystyle\frac{n+1}{2}; & 0 \cdots 0 & 1 \cdots 1 & \displaystyle\frac{n}{2} \end{pmatrix}.$$

Therefore it is of $W$-dimension $2$.

It remains to show that the system cannot be of $C$-dimension $2$.  Suppose on the contrary, it is the intersection of two complete systems $S_1$ and $S_2$.

Consider two winning coalitions of our system: $\{1,\ldots, k+1, n+1,\ldots, n+k+1\}$ and $\{1,\ldots, k, k+2, n+1, \ldots, n+k, n+k+2\}$.  Exchange $k+1$ and $n+k+2$, we have losing coalitions $L_1=\{1,\ldots, k, n+1, \ldots, n+k+2\}$ and $L_2=\{1,\ldots, k+2, n+1, \ldots, n+k\}$.  Since $S_1$ and $S_2$ are swap robust, we can assume that $L_1$ is a winning coalition of $S_1$ but not of $S_2$, and $L_2$ is a winning coalition of $S_2$ but not of $S_1$.

Consider another two winning coalitions:  $\{1,\ldots, k+1, n+1,\ldots, n+k+1\}$ and $\{1,\ldots, k-1, k+1, k+2, n+1, \ldots, n+k, n+k+2\}$. Exchange $k$ and $n+k+2$, we have losing coalitions $L_3=\{1,\ldots, k-1, k+1, n+1, \ldots, n+k+2\}$ and $L_2$. Since $L_2$ is a winning coalition of $S_2$, we have $L_3$ is a winning coalition of $S_1$.

Consider yet another two winning coalitions : $\{1,\ldots, k+1, n+1,\ldots, n+k+1\}$ and $\{1,\ldots, k, k+2, n+1, \ldots, n+k, n+k+3\}$. Exchange $k+1$ and $n+k+2$, we have losing coalitions $L_4=\{1,\ldots, k, n+1, \ldots, n+k+1, n+k+3\}$ and $L_2$.  Once again, $L_4$ is a winning coalition of $S_1.$

Consequently if  $|L\cap \{1,\ldots, n\}| \ge k$ and  $|L\cap \{n+1,\ldots, 2n\}| \ge k+2$,  then $L$ is a winning coalition of $S_1$;  if $|L\cap \{1,\ldots, n\}| \ge k+2$ and  $|L\cap \{n+1,\ldots, 2n\}| \ge k$,  then $L$ is a winning coalition of $S_2$.

Consider $\{1,\ldots, k, k+2, k+3,  n+1, \ldots, n+k, 2n+1\}$ and $\{1,\ldots, k+1, n+1,\ldots, n+k+1\}$.  Exchange $2n+1$ and $k+1$. We get losing coalitions $L_5=\{1,\ldots, k+3, n+1, \ldots, n+k\}$ and $L_6=\{1,\ldots, k, n+1, \ldots, n+k+1, 2n+1\}$.  $L_5$ is a winning coalition of $S_2$ and so $L_6$ is a winning coalition of $S_1$.

Consider $\{1,\ldots, k-1,  n+1, \ldots, n+k+2, 2n+1\}$ and $\{1,\ldots, k+1, n+1,\ldots, n+k+1\}$.  Exchange $2n+1$ and $k+1$. We get losing coalitions $L_7=\{1,\ldots, k-1, k+1, n+1, \ldots, n+k+2\}$ and $L_6$. Both are winning coalitions of $S_1$ and so none are winning coalitions of $S_2$, contradicting that $S_2$ is swap robust.

Thus our system cannot be of $C$-dimension $2$ and therefore has $C$-dimension $3$.
\end{enumerate}
$\hfill\blacksquare$

\section{Power Indices}
In this section, we will try to quantify the power of the Hong Kong government within Legco. Indeed, we are  more interested in the ratio of the power of the Hong Kong government to the power of an elected member of Legco. We will consider the two most common power indices,namely, the Shapley-Shubik Index (see Shapley and Shubik (1954)) and the Bahzhaf Index (see Bahzhaf (1965)).

These two power indices are ordinally equivalent, and therefore rank players equally, for weakly complete games and also for larger classes of cooperative
games like semi-coherent and coherent games (see Freixas (2010)). However, when these two power indices are compared, in the class of weakly complete
games, with other well-known power indices, then the equivalent ranking provided by Shapley-Shubik and Banzhaf indices do not necessarily coincide with the
rankings of some others well-known power indices such as the Johnston index (see Freixas et al. (2012)).

\subsection{Banzhaf Index}

For a member $k \in \{1,...,2n+1\}$, we define $b(k)$ to be the number of winning coalitions $S$ such that  $k\in S$ and $S\backslash\{k\}$ is losing. 
Then the Banzhaf Index of $k$ is defined as
$$\BI(k)=\frac{b(k)}{b(1)+\ldots+b(2n+1)}.$$
For each $1\le j\le 2n$, we have ${2n-1 \choose n}$ such winning coalitions that contain $2n+1$ and $\sum_{r=\lfloor n/2 \rfloor+1}^n {{n-1} \choose {\lfloor n/2 \rfloor}}{n \choose r}$ such winning coalitions that do not contain $2n+1$.  Hence
$$b(j)=\begin{cases} {2n-1 \choose n}+{n-1 \choose {(n-1)/2}}2^{n-1} & n \mbox{ is odd}\\
{2n-1 \choose n}+{n-1 \choose n/2}\left( 2^{n-1} - \frac12{n \choose n/2} \right) & n \mbox{ is even}
\end{cases}$$
For $k=2n+1$, we have
\begin{eqnarray*}
b(2n+1)&=&\sum_{r=n+1}^{2n} {2n \choose r} - \sum_{r=\lfloor n/2 \rfloor+1}^n\sum_{s=\lfloor n/2 \rfloor+1}^n{n \choose r}{n \choose s}\\
&=&\begin{cases} 2^{2n-1}-\frac12{2n \choose n} - (2^{n-1})^2 & n \mbox{ is odd}\\
2^{2n-1}-\frac12{2n \choose n} -\left( 2^{n-1} - \frac12{n \choose n/2} \right)^2 & n \mbox{ is even}
\end{cases}\\
&=&\begin{cases} 2^{2n-2}-\frac12{2n \choose n}& n \mbox{ is odd}\\
2^{2n-2}-\frac12{2n \choose n} + 2^{n-1}{n \choose n/2} -\frac14{n \choose n/2}^2 & n \mbox{ is even}
\end{cases}
\end{eqnarray*}

Since $\BI(j)=\BI(1)$ for all $j$ ($1 \le j \le 2n$), we shall only consider the ratio of $\BI(2n+1)$ to $\BI(1)$ which is equal to $\frac{b(2n+1)}{b(1)}$.

\begin{figure} \label{fig2}
\includegraphics[scale=0.6]{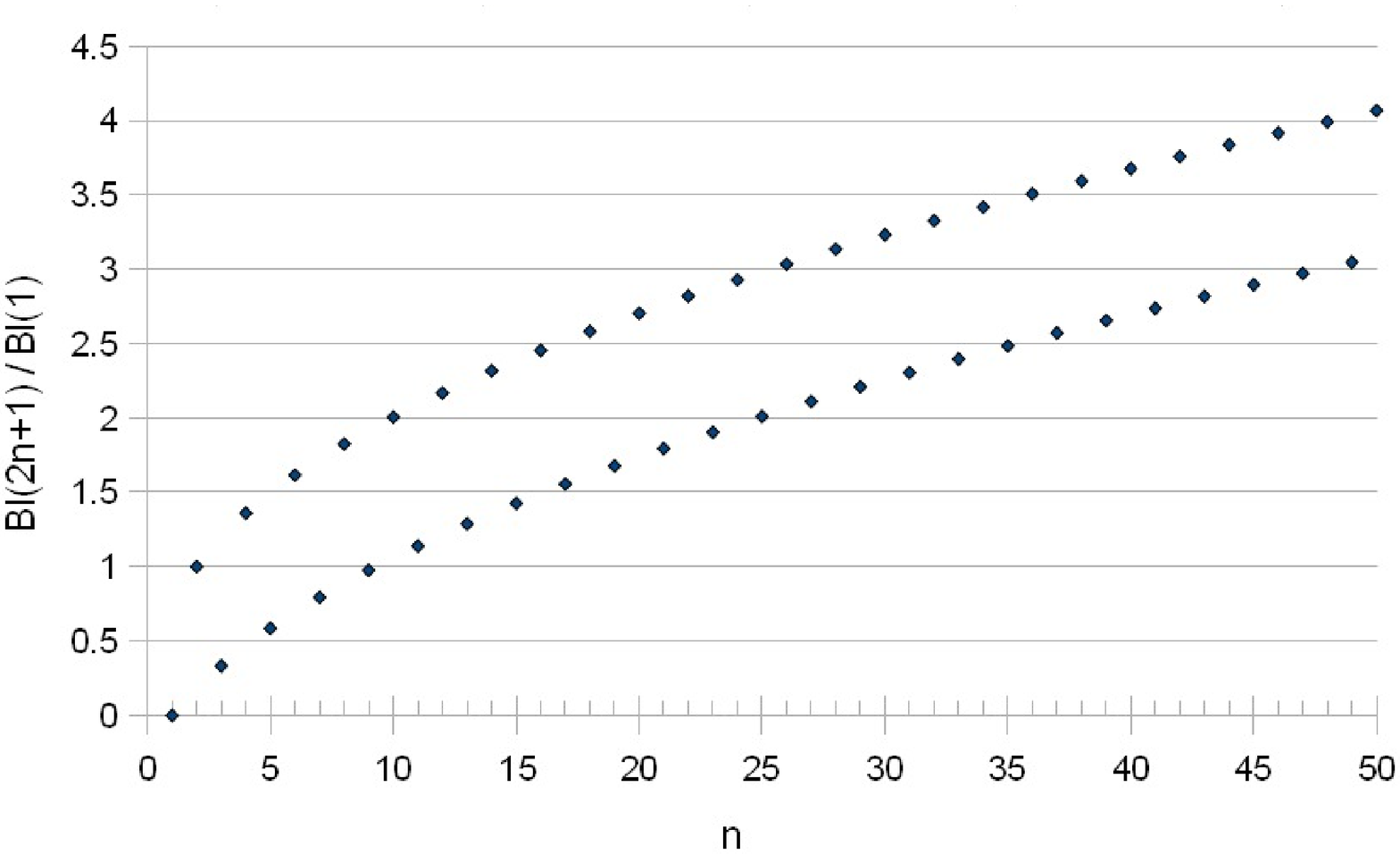}
\caption{The ratio $\BI(2n+1)/\BI(1)$}
\end{figure}

As seen in both Figure 1, the ratio of the power of $2n+1$ to that of any $j$ ($1\le j\le 2n$) increases in general as $n$ increases, but the ratio when $n$ is odd is well below when $n$ is even.  
Heuristically, it is likely due to the fact that the power of $2n+1$ is cardinally greater than that of $k$ (i.e. $2n+1 \ge_D k$) when $n$ is even, but this is not the case when $n$ is odd, as shown in Proposition $1$. To prove this observation for the Banzhaf Indice in Figure $1$ mathematically, we apply the Stirling's approximation for the binomial coefficients: ${n \choose k}\approx \frac{(\frac{n}{k}-\frac12)^ke^k}{\sqrt{2\pi k}}$. When $n$ is odd and large, we then have $$\frac{\BI(2n+1)}{\BI(1)}=\frac{b(2n+1)}{b(1)}\approx \sqrt{\pi/2}\sqrt{n}(\sqrt{2}-1)-\sqrt{2}(\sqrt{2}-1).$$ While for $n$ is even and large, we have  $$\frac{\BI(2n+1)}{\BI(1)}=\frac{b(2n+1)}{b(1)}\approx \sqrt{\pi/2}\sqrt{n}(\sqrt{2}-1)+ \sqrt{2}-1.$$ Thus, $\frac{\BI(2n+1)}{\BI(1)}=O(\sqrt{n})$ and the gap between the two curves in Figure 1 is about $\sqrt{2}-1 +\sqrt{2}(\sqrt{2}-1)=1$.

\subsection{Shapley-Shubik Index}

\begin{figure} \label{fig1}
\includegraphics[scale=0.6]{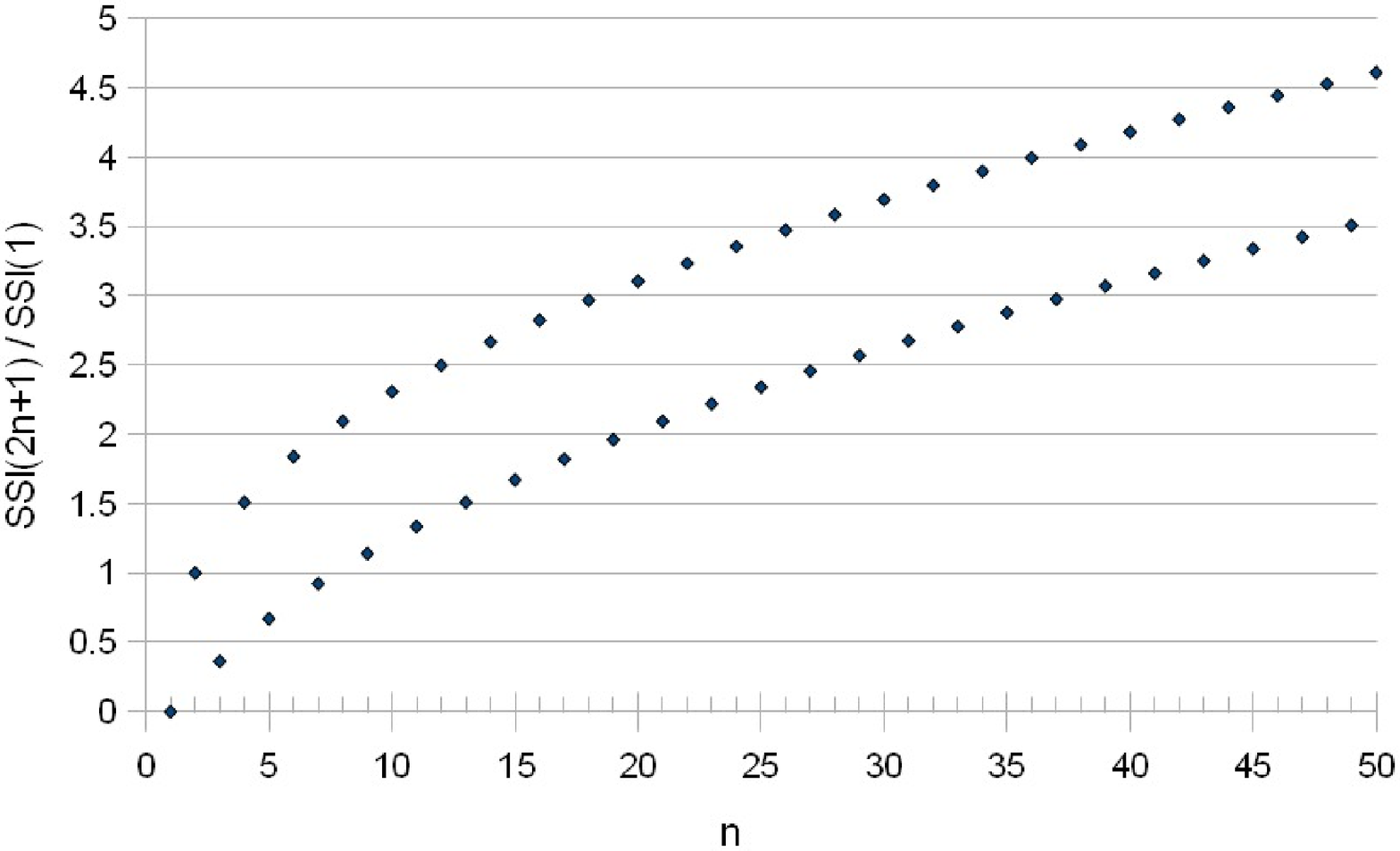}
\caption{The ratio $\SSI(2n+1)/\SSI(1)$}
\end{figure}

Let $\pi=\begin{pmatrix} 1 &\cdots &n \\ \pi_1 &\cdots &\pi_n \end{pmatrix}$ be a permutation on $\{1,\ldots,n\}$, then there exists a unique $j$ such that $\{\pi_1,\ldots,\pi_{j-1}\}$ is not a winning coalition but $\{\pi_1,\ldots,\pi_j\}$ is a winning coalition, and we write $v_\pi=\pi_j$.

For any $k\in\{1,\ldots,n\}$, the Shapley-Shubik Index of a member $k$ is defined as
$$\SSI(k)=\frac 1{n!} |\{\pi\ :\ v_\pi=k\}|.$$ 
Note that $\sum_{j=1}^n \SSI(j)=1$.

Now consider the invisible representative $2n+1$ in Legco and the permutation 
$$\pi=\begin{pmatrix} 1 &\cdots &k& k+1& k+2& \cdots &2n+1 \\ \pi_1 &\cdots &\pi_k&2n+1&\pi_{k+1}&\cdots&\pi_{2n} \end{pmatrix}.$$

Let $r=|\{\pi_1,\ldots,\pi_k\}\cap\{1,\ldots,n\}|$ and $s=k-r=|\{\pi_1,\ldots,\pi_k\}\cap\{n+1,\ldots,2n\}|$.  

Then $v_\pi=2n+1$ if and only if $r+s=k\ge n+1$ but one of $s$ and $r$ is less than $\lfloor n/2 \rfloor+1$, i.e., either
\begin{enumerate}
\item[(i)] $1\le r\le \lfloor n/2 \rfloor$ and $n+1-r\le s\le n$; or
\item[(ii)] $1\le s\le \lfloor n/2 \rfloor$ and $n+1-s\le r\le n$.
\end{enumerate}
Note that they are two disjoint cases and have the same number of permutations.
Therefore
\begin{eqnarray*}
\SSI(2n+1)&=&2\times \frac{1}{(2n+1)!}\sum_{r=1}^{\lfloor n/2 \rfloor}\sum_{s=n+1-r}^{ n}{n \choose r}{n \choose s}(r+s)!(2n-r-s)!\\
&=&\frac{2}{2n+1}\left(\sum_{r=1}^{\lfloor n/2 \rfloor}\sum_{s=n+1-r}^{n}{r+s \choose r}{2n-r-s \choose n-r}\right) {{2n} \choose {n}}^{-1}\\
\end{eqnarray*}


We do not find any further simplifications of the last expression but numerical data leads us to make the following 

\noindent
{\bf Conjecture 1:} $\SSI(2n+1) \sim \lambda n^{-1/2}$ for some positive constant $\lambda$ when $n$ tends to infinity (which is equivalent to saying that the double sum in $\SSI(2n+1)$ above is approximately $\lambda 4^n$).

\noindent
{\bf Conjecture 2:}  $\SSI(2n+1)/\SSI(1)=O(\sqrt{n})$. 

\medskip
Note that since  $\SSI(j)=\SSI(1)$ (as $j=_D1$) for all $1 \le j \le 2n$ and $\sum_{j=1}^{2n} \SSI(j)+\SSI(2n+1)=1$,  
the ratio of $\SSI(2n+1)$ to $\SSI(1)$ equals to $\frac{2n\SSI(2n+1)}{1-\SSI(2n+1)}$. Conjecture 1 will then imply 
$\frac{\SSI(2n+1)}{\SSI(1)}\sim 2\lambda\sqrt{n}$ and hence Conjecture 2.

If Conjecture 2 is true, then we have both $\frac{\SSI(2n+1)}{\SSI(1)}$ and $\frac{\BI(2n+1)}{\BI(1)}$ are of same order of growth $O(\sqrt{n})$. Moreover, comparing Figure 1 and Figure 2, it seems that both $\frac{\SSI(2n+1)}{\SSI(1)}$ and $\frac{\BI(2n+1)}{\BI(1)}$ share the same asymptotic gap behavior. Unfortunately, we cannot find an explanation of this gap behavior for the Shapley-Shubik index. 

%
%

Finally, as one can see from Figure $2$, for the current Legco ($n=35$), the power of the Hong Kong government is almost three times that of any member in Legco as $\SSI(71)=0.0395$ and $\SSI(k)=0.0137$ for $k \le 70$.


\section{Conclusion}

The pace of constitutional reform in Hong Kong is a central issue during the five-month public consultation exercise for the {\it``Methods for Selecting the Chief Executive in 2017 and for Forming the Legislative Council in 2016 Public Consultation''}. The  mathematical model of Legco introduced in section 3 allows one to quantify the pace of constitutional reform of Legco from the point of view of its complexity. In fact, abolishing the difference between a bill proposed by the government and a Legco member, while keeping the functional constituencies and the split-voting mechanism will make Legco a bicameral system and its dimension will be reduced from three to two. If one further abolishes the split-voting mechanism, the dimension will then be reduced from two to one (even one keeps the functional constituencies). This shows that it is the role of the government and the presence of the split-voting mechanism that determine the dimension of Legco, not the existence of functional constituencies. In addition, we would also like to point out that if one only broadens the electorate base of the functional constituencies and implement the so-called "one person two votes" (i.e., each voter has one vote in the direct geographical election and has one vote in one of the functional constituencies), the dimension of Legco will still be three and therefore, there is no change in the complexity of the voting system of Legco. 

Another possible approach to the constitutional reform in Hong Kong is to increase the number of the members of both the geographical constituencies and the functional constituencies (i.e., to increase $n$ in Legco) while keeping the split-voting mechanism. In section 4, by computing two common power indices, we find out that the relative power of the Hong Kong government to any member of Legco increases as odd $n$ increases or even $n$ increases. So according to our model, the Hong Kong government can actually gain more power (compared with any individual member) if the size of Legco is increased from $70$ to $2n$ for any even $n > 35$.   

\section*{Acknowledgment}
The authors want to thank the referee for his/her many very helpful suggestions, especially, the suggestion of adding a discussion of $C$-dimension and $W$-dimension of Legco and the use of Stirling's approximations for comparing $B(2n+1)$ with $B(1)$. We also include some of his/her very valuable comments in this paper.

\end{document}